\documentclass{article}
\usepackage{amssymb}

\input{tcilatex}
\begin{document}

\begin{center}
{\LARGE Introduction to the multiple-quantum}

{\LARGE operator spaces\vspace{0.05in}\vspace{0.1in}}\\[0pt]
{\large Xijia Miao}

{\large Somerville, Massachusetts\\[0pt]
\vspace{0.02in}Date: June 2009}
\end{center}

The multiple-quantum NMR spectroscopy has an extensive application in
determination of the bio-macro-molecular structures and in the investigation
of the properties of a variety of physical materials. In quantum computation
the multiple-quantum transition processes have been used to construct the
quantum circuits, quantum algorithms, and quantum simulations. The
multiple-quantum operator algebra spaces are closely related to the
symmetries of a multiple-spin quantum system. They may have an important
effect on the multiple-quantum transition processes and the multiple-quantum
NMR spectroscopy of the spin system. Here gives a brief introduction to the
multiple-quantum operator algebra spaces.\newline
\newline
{\large 1. The multiple-quantum operator algebra spaces}

Symmetry in quantum mechanics is important. Symmetry of the Hilbert state
space of a quantum system is one of the most important factors to help the
unitary quantum dynamics to speed up a quantum computation [1, 2]. Symmetry
leads to that the whole Hilbert space of a quantum system or equivalently
the whole Liouville operator space of a quantum ensemble is decomposed into
many small subspaces. This further simplifies the unitary quantum dynamics
of the quantum system. Thus, people may use some symmetric properties of a
quantum system to simplify investigation of the unitary time evolution
process of the system. For example, the rotating symmetry in the spin
internal space results in that the total spin angular momentum of a spin
system is constant [3, 4]. Symmetry is closely related to degeneracy of
energy eigenvalues in a quantum system. In an uncoupled or weakly coupled
multiple-spin system (there is not any orbital angular momentum) it is
usually related to degeneracy of eigenvalues of the total spin angular
momentum as the total spin angular momentum commutes with the spin
Hamiltonian of the system. There is also another important symmetry, i.e.,
the interchange symmetry between a pair of spins in an uncoupled
multiple-spin system [3, 4, 5]. This interchange symmetry leads to that
eigenvalues of the total spin angular momentum and its z-component are
degenerate in an uncoupled multiple-spin system. The multiple-quantum
operator algebra spaces [6] that will be briefly introduced below are
closely related to the interchange symmetry in a multiple-spin system.

According to quantum mechanics [3] a quantum system may be described by a
wave function. There is a Hilbert state space for the quantum system and
each quantum state to describe the quantum system belongs to the Hilbert
space. A dynamical variable then is represented by a linear operator in
quantum mechanics. On the other hand, a quantum system which may be a
pure-state system or a quantum ensemble also may be described by a density
operator. This density-operator-based description has been used extensively
in the NMR spectroscopy [7] and a number of other research fields, whose
research object usually is a quantum ensemble. In quantum mechanics both the
descriptions based on a wave function and on a density operator are
compatible with the fundamental and universal principle that both a closed
quantum system and its ensemble obey the same unitary quantum dynamics [1,
2, 6, 9, 11]. A spin system usually has a finite-dimensional Hilbert space.
To describe generally the time evolution process of a quantum system with a
finite-dimensional Hilbert space it could be more convenient to use a
complete set of the base operators to expand the density operator of the
system [8]. Such a complete set of the base operators form an operator space
called the Liouville operator space [7]. This Liouville operator space
(sometime it is called the Hilbert operator space) corresponds to the
Hilbert state space of the same quantum system. Obviously, any dynamical
variable such as the Hamiltonian of the quantum system also may be expanded
in terms of the base operators. Therefore, the symmetric structure and
property of the Liouville operator space may be helpful for simplifying
investigation of the time evolution process of a quantum system. For a
multiple-spin system\ both the rotating symmetry in the spin internal space
and the interchange symmetry could be exploited to help the investigation.
Due to the two symmetries and especially the interchange symmetry of a
multiple-spin system the Liouville operator space of an $n-$spin system may
be generally divided into several operator subspaces [6]:

(i) The longitudinal magnetization and spin order ($LOMSO$) operator subspace

(ii) The zero-quantum operator subspace

(iii) The even-order multiple-quantum operator subspace\newline
Here the largest subspace is the even-order multiple-quantum operator
subspace which contains the first two subspaces. The second largest one is
the zero-quantum operator subspace which contains the whole $LOMSO$ operator
subspace. The $LOMSO$ operator subspace is the minimum subspace among the
above three subspaces. This is a commutable subspace. It usually consists of
all the diagonal operators of the Liouville operator space of a quantum
system.

The usual observation in the multiple-quantum NMR spectroscopy is not
correct that all zero-quantum coherence operators of a spin system form a
closed operator subspace or that all zero-quantum coherence operators plus
the longitudinal magnetization operators of a spin system form a closed
operator subspace. A closed zero-quantum operator subspace must contain the
whole $LOMSO$ operator subspace, that is, it consists of all the
zero-quantum coherence operators, longitudinal magnetization operators,
longitudinal spin order operators, and the unity operator. Here the unity
operator is a necessary component to form a closed operator algebra space.

Furthermore, each one of these three operator subspaces could be further
divided into several smaller operator subspaces. In particular, the above
zero-quantum operator subspace may be further divided into the $(n+1)$
selective zero-quantum operator subspaces for an $n-$spin $(I=1/2)$ system
[9], each one of which is much smaller than the total zero-quantum operator
subspace.

The conventional zero- and double-quantum operator subspaces [7] in a
two-spin $(I=1/2)$ system are very special. They are deduced based on the
corresponding relations between them and the operator space of a single spin
(I=1/2) system. It is known that the operator space of a single-spin (I=1/2)
system is given by $\{I_{x},$ $I_{y},$ $I_{z}\}.$ It can turn out that for
the operator commutation relations there exist the corresponding relations: 
\[
\frac{1}{2}(I_{1z}-I_{2z})\leftrightarrow I_{z};\text{ }%
(I_{1x}I_{2x}+I_{1y}I_{2y})\leftrightarrow I_{y};\text{ }%
(I_{1x}I_{2y}-I_{1y}I_{2x})\leftrightarrow I_{x}. 
\]%
It can be deduced from these corresponding relations that these three
operators $\{(I_{1x}I_{2y}-I_{1y}I_{2x}),$ $(I_{1x}I_{2x}+I_{1y}I_{2y}),$ $%
\frac{1}{2}(I_{1z}-I_{2z})\}$ form an operator space. This operator space is
just the conventional zero-quantum operator space of a two-spin ($I=1/2$)
system. Thus, the usual zero-quantum operator subspace of a two-spin system
and the operator space of a single-spin system are isomorphic to each other.
While one may use these corresponding relations to deduce the above
zero-quantum subspace of a two-spin system, the subspace is very special.
These corresponding relations can not be used to further deduce a general
zero-quantum operator subspace in a complex spin system. Similarly, the
conventional double-quantum operator subspace in a two-spin system is also
very special. The corresponding relations similar to the above can not be
used to further deduce a general even-order multiple-quantum operator
subspace in a complex spin system.

The multiple-quantum operator spaces may be used to simplify the quantum
simulation of a general spin quantum system. Consider the spin Hamiltonian
of an $n-$qubit spin system. The spin Hamiltonian could not be in any one of
the above three operator subspaces, but it is still in the Liouville
operator space (here without considering the normalization condition), that
is, the spin Hamiltonian still can be expanded in terms of the complete base
operator set of the Liouville operator space. The unitary propagator of the
spin Hamiltonian can be decomposed into a sequence of many elementary
propagators. Obviously, complexity of the decomposition may be measured by
the number of elementary propagators in the decomposed sequence. Then the
multiple-quantum operator spaces may be used to decrease the number of
elementary propagators in the decomposed sequence [6]:

(i) At first the unitary propagator (or the spin Hamiltonian) is converted
into an element of the even-order multiple-quantum operator subspace by a
unitary transformation

(ii) The ever-order multiple-quantum unitary propagator then is converted
into an element of the zero-quantum operator subspace by the ever-order
multiple-quantum unitary transformation

(iii) The zero-quantum unitary propagator finally is converted into an
element of the $LOMSO$ operator subspace by a zero-quantum unitary
transformation.

One of the important properties of an operator algebra space is the closure
property. This closure property leads directly to the fact that in a
zero-quantum operator subspace the product of any two zero-quantum operators
is still a zero-quantum operator. This further leads to the fact that a
unitary exponential operator generated by a hermite zero-quantum coherence
operator is also a zero-quantum operator. Suppose that $Z_{q}$ is a hermite
zero-quantum coherence operator. Then the unitary operator generated by the
zero-quantum operator $Z_{q}$ is written as%
\begin{equation}
U_{zq}(t)=\exp \{-iZ_{q}t\}.  \tag{1}
\end{equation}%
Then it can turn out that this unitary operator $U_{zq}(t)$ is also a
zero-quantum operator and is called the zero-quantum unitary operator.
Denote $\{Z_{a}\}$ as the base operator set of the zero-quantum operator
subspace. With the help of the closure property of a zero-quantum operator
subspace one can prove that the following zero-quantum unitary
transformation may be expanded in the zero-quantum operator subspace:%
\begin{equation}
U_{zq}(t)Q_{zq}U_{zq}(t)^{-1}=\sum_{a}A_{a}Z_{a},  \tag{2}
\end{equation}%
where $Q_{zq}$ is a traceless and hermite zero-quantum operator, while the
zero-quantum base operators $\{Z_{a}\}$ include the longitudinal
magnetization operators $\{I_{kz}\}$ and spin order operators $%
\{2I_{kz}I_{lz},$ $4I_{kz}I_{lz}I_{mz},$ $...\}$ as well as the zero-quantum
coherence operators $\{(ZQC)_{k}\}$. Then the unitary transformation may be
explicitly written as%
\[
U_{zq}(t)Q_{zq}U_{zq}(t)^{-1}=\sum_{k=1}^{n}\Omega
_{k}(t)I_{kz}+\sum_{k>l}J_{kl}(t)2I_{kz}I_{lz} 
\]%
\begin{equation}
+\sum_{k>l>m}J_{klm}(t)4I_{kz}I_{lz}I_{mz}+...+J_{12...n}(t)2^{n-1}I_{1z}I_{2z}...I_{nz}+\sum_{k}\beta _{k}(t)(ZQC)_{k},
\tag{3}
\end{equation}%
where $\Omega _{k}(t),$ $J_{kl}(t),$ $J_{klm}(t),$ $J_{12...n}(t),$ $\beta
_{k}(t),$ $etc.,$ are the amplitude parameters. This formula shows that if
the zero-quantum operator $Q_{zq}$ is a longitudinal magnetization operator,
e.g., $I_{kz},$ then it can be converted into not only the longitudinal
magnetization operators $\{I_{lz}\}$ but also many spin-order operators such
as $2I_{kz}I_{lz},$ $4I_{kz}I_{lz}I_{mz},$ $etc.,$ as well as the
zero-quantum coherence operators $\{(ZQC)_{k}\}$. This formula could be
useful to analyze in theory some NMR experiments. Consider a conventional
nuclear spin diffusion experiment in the NMR spectroscopy [7]. Here assume
that the spin diffusion is driven by a zero-quantum Hamiltonian and the
initial density operator of the spin diffusion process is taken as the
longitudinal magnetization such as $I_{kz}.$ Then during the spin diffusion
process the initial longitudinal magnetization $I_{kz}$ is converted into
other longitudinal magnetizations $\{I_{lz}\}$ with $l\neq k.$ This is the
desired result for the spin diffusion experiment. However, the formula (3)
shows that the initial longitudinal magnetization also may be converted into
the longitudinal spin orders and the zero-quantum coherences. These terms
are undesired for the spin diffusion experiment and need to be cancelled in
experiment. Therefore, during the NMR detection of the spin diffusion one
usually needs to use a purging pulse (or the homonuclear decoupling
techniques) to cancel these undesired terms.

Note that the total zero-quantum subspace may be further divided into the $%
(n+1)$ selective zero-quantum subspaces for an $n-$spin (I=1/2) system. Then
the above zero-quantum unitary transformation even can be greatly
simplified. Denote $S(k)\times S(k)$ as the $k-$th selective zero-quantum
operator subspace for $k=0,1,...,n$ [9]. Now the hermite zero-quantum
operator $Z_{q}$ in Eq. (1) may be projected onto these $(n+1)$ zero-quantum
subspaces,%
\begin{equation}
Z_{q}=\sum_{k=0}^{n}Z_{q}(k).  \tag{4}
\end{equation}%
Since each one of these $(n+1)$ zero-quantum operators $\{Z_{q}(k)\}$
belongs to a different zero-quantum subspace, all these $(n+1)$ zero-quantum
operators commute with each other. Moreover, the $k-$th zero-quantum
operator $Z_{q}(k)$ can affect only those operators of the $k-$th
zero-quantum subspace $S(k)\times S(k).$ These results show that if the
zero-quantum operator $Q_{zq}^{k}$ belongs to the $k-$th zero-quantum
subspace $S(k)\times S(k),$ then the unitary transformation $%
U_{zq}(t)Q_{zq}^{k}U_{zq}(t)^{-1}$ also generates a zero-quantum operator of
the same subspace $S(k)\times S(k)$ and it may be expanded as 
\begin{equation}
U_{zq}(t)Q_{zq}^{k}U_{zq}(t)^{-1}=\sum_{a}A_{a}^{k}Z_{a}(k),  \tag{5}
\end{equation}%
where $\{Z_{a}(k)\}$ are the base operator set of the zero-quantum subspace $%
S(k)\times S(k).$ Note that $Z_{q}$ can be an arbitrary hermite zero-quantum
operator. The formula (5) shows that the zero-quantum operator $Q_{zq}^{k}$
evolves independently in its own selective zero-quantum subspace $S(k)\times
S(k)$ under the unitary transformation and any two selective zero-quantum
subspaces do not mix with other each during the unitary transformation. This
point is particularly important to analyze or design some NMR experiments
such as the spin diffusion experiment. In general, the initial density
operator of the spin diffusion process may not be in some selective
zero-quantum subspace. But it can be projected onto these selective
zero-quantum subspaces. Then one may use the formula (5) to calculate
independently the time evolution process of the spin diffusion in each one
of these selective zero-quantum subspaces. Therefore, the formula (5) may
simplify the theoretical calculation of the time evolution process of the
nuclear spin diffusion.

It is known that each one of these $(n+1)$ selective zero-quantum subspaces $%
\{S(k)\times S(k)\}$ for $k=0,1,...,n$ has a different dimensional size $%
d(k)^{2}$ with the number $d(k)$ given by [10, 9]%
\begin{equation}
d(k)=\frac{n!}{k!(n-k)!}.  \tag{6}
\end{equation}%
For $k\thicksim n/2$ these zero-quantum subspaces $\{S(k)\times S(k)\}$ have
exponentially large dimensions. If the spin diffusion occurs in these
exponentially large zero-quantum subspaces, then thing becomes much
complicated in the theoretical treatment. On the other hand, it becomes much
simpler to calculate the time evolution process of the spin diffusion if the
spin diffusion is confined in the polynomially small zero-quantum subspaces
or occurs mainly in these polynomially small subspaces. From the point of
view of the NMR techniques it could not be difficult to prepare a spin
system in some polynormially small zero-quantum subspaces in the spin
diffusion experiment. The problem is how to measure efficiently the NMR
signals from different zero-quantum subspaces.

The above theoretical strategy is not new that one uses the small
zero-quantum subspaces to realize the NMR parameter measurement (or to
calculate the corresponding time evolution process) in the nuclear spin
diffusion experiment. In methodology it is really equivalent to the one that
one uses small state subspaces of the Hilbert space to realize a quantum
computation (or to calculate the corresponding time evolution process) in a
pure-state quantum system. This strategy has been used to help to solve
efficiently the quantum search problem [1, 2, 9] in quantum computation.
Since the strategy may work in a pure-state quantum system and also in a
quantum ensemble, the fundamental and universal quantum-mechanical principle
that both a closed quantum system and its quantum ensemble obey the same
unitary quantum dynamics [1, 2, 6, 9, 11] is of crucial importance when the
strategy is used in quantum computation. This strategy might also be used to
realize other quantum algorithms based on the unitary quantum dynamics and
the quantum simulations. Here it will not be discussed in detail. \newline
\newline
{\large 2. Some important properties of a zero-quantum coherence operator}

A zero-quantum coherence operator also has some important properties which
are independent of the multiple-quantum operator algebra spaces mentioned
above. One of these important properties is stated as follows: any $p-$order
quantum coherence operator does not change its quantum coherence order when
it is acted on by a zero-quantum coherence operator. This property was first
described qualitatively in the multiple-quantum NMR spectroscopy [12, 7].
Later, unknowing this description, the present author proved rigorously this
property in a general spin system on the basis of the operator algebra
theory [13]. With the help of the property of a zero-quantum coherence
operator one can realize that the product operator formalism [14] can be
used to analyze exactly the time evolution process in a strongly coupled
spin system just like in a weakly coupled spin system [13]. The property
also could be useful in quantum computation. However, it is necessary to
have a strict mathematical proof for this property in a general spin system
before the property can be used generally in the research fields such as the
quantum simulation in quantum computation. Thus, only when these works [12,
7, 13] relevant to the property are put together, can the description for
the property be considered to be complete.

Another important property of a zero-quantum coherence operator is that
eigenvalues of any zero-quantum coherence operator are zero for both the
ground state and the highest excited state in an $n-$spin ($I=1/2$) system. 
\newline
\newline
{\large 3. The permutation transformation between the two different
encodings for the product bases}

The conventional computational basis set of an $n-$qubit spin system has
been used extensively in quantum computation. This basis set is really the
conventional product basis set of the $n-$spin ($I=1/2$) system that has
been used extensively in the NMR spectroscopy [10, 13]. Obviously, each
product base is an eigenstate of the total longitudinal magnetization
operator $I_{z}=\sum_{k=1}^{n}I_{kz}$ of the $n-$spin (I=1/2) system.
However, almost every eigenvalue of the longitudinal magnetization operator $%
I_{z}$ is degenerate in the spin system. The longitudinal magnetization
operator has a diagonal representation matrix in these product bases. Now
these product bases are arranged suitably such that the diagonal matrix
element, which is really an eigenvalue of the operator $I_{z},$ descends (or
does not increase) from top to bottom. Such an encoding for the product
bases is more convenient to study the multiple-quantum transition processes
[9]. It is different from the encoding of the conventional computational
bases. Obviously, there is a permutation transformation between the two
encodings of the product bases. It can turn out that this permutation
transformation may be realized efficiently by the unitary operations
generated by the anti-diagonal hermite matrices [9]. \newline
\newline
{\large Note added }

It is generally hard to calculate exactly the time evolution process of the
nuclear spin diffusion in solid in the NMR spectroscopy. The NMR\
Researchers tend to use a variety of approximation methods to treat in
theory the nuclear spin diffusion, as can be seen in several works reported
in the recent NMR meetings. The multiple-quantum operator spaces could
provide a general theoretical frame to treat and explain the nuclear spin
diffusion. \newline
\newline
{\large References}\newline
1. X. Miao, \textit{Universal construction for the unsorted quantum search
algorithms}, http://arxiv.org/abs/quant-ph/0101126 (2001); \textit{Solving
the quantum search problem in polynomial time on an NMR quantum computer},
http://arxiv. org/abs/quant-ph/0206102 (2002); \textit{Quantum search
processes in the cyclic group state spaces}, http://
arxiv.org/abs/quant-ph/0507236 (2005); \newline
2. X. Miao, unpublished\newline
3. M. E. Rose, \textit{Elementary theory of angular momentum}, Wiley, New
York, 1957\newline
4. A. R. Edmonds, \textit{Angular momentum in quantum mechanics}, 2nd edn,
Princeton University Press, Princeton, 1974\newline
5. L. I. Schiff, \textit{Quantum mechanics}, 3rd, McGraw-Hill book company,
New York, 1968\newline
6. X. Miao, Mol. Phys. 98, 625 (2000)\newline
7. R. R. Ernst, G. Bodenhausen, and A. Wokaun, \textit{Principles of nuclear
magnetic resonance in one and two dimensions}, Oxford University Press,
Oxford, 1987\newline
8. U. Fano, Rev. Mod. Phys. 29, 74 (1957)\newline
9. X. Miao, \textit{the efficient multiple-quantum transition processes in
an n-qubit spin system}, http://arxiv.org/abs/quant-ph/0411046 (2004)\newline
10. P. L. Corio, \textit{Structure of high-resolution NMR spectra}, Academic
Press, New York, 1966\newline
11. X. Miao, \textit{A polynomial-time solution to the parity problem on an
NMR quantum computer}, http://arxiv.org/abs/quant-ph/0108116 (2001); \textit{%
The basic principles to construct a generalized state-locking pulse field
and simulate efficiently the reversible and unitary halting protocol of a
universal quantum computer}, http://arxiv.org/abs/quant-ph/0607144 (2006)%
\newline
12. (a) W. S. Warren, D. P. Weitekamp, and A. Pines, J. Chem. Phys. 73, 2084
(1980); (b) G. Drobny, A. Pines, S. Sinto, D. Weitekamp, and D. Wemmer,
Faraday Div. Chem. Soc. Symp. 13, 49 (1979)\newline
13. X. Miao, X. Han, and J. Hu, Science in China A 36, 1199 (1993)\newline
14. (a) O. W. S$\phi $rensen, G. W. Eich, M. H. Levitt, G. Bodenhausen, and
R. R. Ernst, Prog. NMR spectroscopy, 16, 163 (1983); (b) K. J. Packer and K.
M. Wright, Mol. Phys. 50, 797 (1983); (c) F. J. M. Van De Ven and C. W.
Hilbers, J. Magn. Reson. 54, 512 (1983)\newline
\newline

\end{document}